# Rogue waves excitation on zero-background in the (2+1)-dimensional KdV equation

Jie-Fang Zhang[1*], Mei-zhen Jin[2], and Meng-yang Zhang[3]

a) Institute of Intelligent Media Technology, Communication University of Zhejiang, Hangzhou 310018, Zhejiang, China;

b) Network and Data Center, Communication University of Zhejiang, Hangzhou 310018, Zhejiang, China;

c) Covestro (Shanghai) Investment Company Limited, Shanghai 201507, China.

**Abstract:** An analytical method for constructing various coherent localized solutions with short-lived characteristics is proposed based on a novel self-mapping transformation of the (2+1) dimensional KdV equation. The highlight of this method is that it allows one to generate a class of basic two--dimensional rogue waves excited on zero-background for this equation, which includes the line-soliton-induced rogue wave and dromion-induced rogue wave with exponentially decaying as well as the lump-induced rogue wave with algebraically decaying in the $(x,y)$-plane. Our finding provides a proper candidate to describe two-dimensional rogue waves and paves a feasible path for studying rogue waves.

**Keywords:** Rogue wave; line-soliton; dromion; lump; (2+1)-dimensional KdV equation

**PACS numbers:** 05.45.Yv, 02.30.Jr, 0340.Kf

---

∗ Corresponding author. Email: zhangjief@cuz.edu.cn

Rogue wave (RW) are a typical natural phenomenon that can occur in a variety of different environments[1-3]. They is displaying the characteristics of large amplitude and "coming without shadow and going without trace". Peregrine first discovered a novel rational solution are localized in both space and time, which is also called 'Peregrine soliton (PS)' [4], in the nonlinear Schrödinger (NLS) equation. Subsequently, the higher order rogue waves (RWs) and multiple RWs of the NLS equation had been found [5-9]. Thus, it has gradually been regarded as the best prototype of RWs and has become a consensus. And it is natural to think that the NLS equation is an ideal model to excite RWs. In the past 20 years, the research on RWs of the (1+1) dimensional nonlinear models has achieved rich results, which can be said to be quite complete.

It is well known that the (2+1)-dimensional nonlinear wave model is a natural extension of the (1+1)-dimension, which is a physical model that is closer to reality than the (1+1) dimensional nonlinear model. The various (2+1)-dimensional nonlinear wave model have been discovered in the natural and applied sciences. Moreover, more abundant localized structures, such as line-solitons, breathers, dromions, lumps, peakons, compactons, foldons as well as the interaction properties, for the (2+1) dimensional integrable models had been revealed. Obviously, the RWs in the (2+1) dimensional nonlinear models has also become a natural scientific problem.Yang and Ohta found the fundamental line RW, which excited on constant background, for both the Davey-Stewartson I(DSI)[10 ] and the Davey-Stewartson II (DSII) [11] equation by the bilinear method, respectively. He et al. [12] further gave doubly localized two-dimensional RWs in the DSI equation by the bilinear method. It is worth pointing out that the resulting multi-order and higher-order RWs of the DS equation which studied in the above literature, are obtained by the interaction between different line-solitons. This is basically similar to the earlier study on the formation of RWs in the KP equation [13,14]. More recently, He et al. [15] investigated the RWs excited on the zero-background, which are localized in both space and time, in a Benney-Roskes model by the even fold Darboux transformation. And very interesting, they found such RWs not only decay quickly and but also change in structure with time, so and could be viewed as a two-dimensional analog of the PS in NLS equation.

Nevertheless, we think that the studying progress of the RWs in the (2+1)-dimensional nonlinear models is not as good as ones in the (1+1) dimensional nonlinear models. Particularly, the RWs excited on zero-background in both one-spatial and two-spatial nonlinear models are still basically open. Motivated by this reason, in this letters, we will start to investigate the RW excited on zero-background of the (2+1)-dimensional KdV equation as a first example.

In analogy with one-dimensional (1D) RWs based on the (1+1) dimensional nonlinear model, the two-dimensional (2D) RWs based on the (2+1) dimensional nonlinear model, are also defined as a class of the short-lived large-amplitude waves, which is doubly localized in two spatial variables $x$ and $y$ as well as in time $t$. But we will generalize their modulus can be both a rational function and an exponential function. This is a bold conjecture that, as far as we know, does not seem to to have been reported in the literature.

We consider the (2+1)-dimensional KdV equation in the following form

$$u_t + u_{xxx} - 3(uv)_x = 0, \quad u_x = v_y, \tag{1}$$

where $u = u(x,y,t)$ represents the physical field and $v = v(x,y,t)$ some potential. Equation (1) reduces to the (1+1)-dimensional KdV equation by setting $v = u$ and $x = y$. Equation (1) was derived by Boiti et al. [16] using the idea of the weak Lax pair.

To investigate the rogue wave aspects of Eq.(1), we shall introduce the following two-dimensional self-mapping transformation (SMT):

$$\begin{aligned} u(x,y,t) &= \rho_1(t)U(\xi(x,t),\eta(y,t),\tau(t)) + \alpha_1(x,y,t), \\ v(x,y,t) &= \rho_2(t)U(\xi(x,t),\eta(y,t),\tau(t)) + \alpha_2(x,y,t), \end{aligned} \tag{2}$$

Our goal is to find out this SMT (2) of the (2+1)-dimensional KdV equation (1), under which the form of (1) remains unchanged, that is (1) is converted into

$$U_\tau + U_{\xi\xi\xi} - 3(UV)_\xi = 0, \quad U_\xi = V_\eta, \tag{3}$$

where $U = U(\xi,\eta,\tau), V = V(\xi,\eta,\tau), \rho_1 = \rho_1(t), \rho_2 = \rho_2(t), \xi = \xi(x,t), \eta = \eta(y,t), \tau = \tau(t),$ $\alpha_1 = \alpha_1(x,y,t), \alpha_2 = \alpha_2(x,y,t)$, which may be obtained by substituting (2) into (1). $\xi, \eta, \tau$ are the self-similar variables between (1) and (2), $\rho_1, \rho_2$ are call as the amplitude scaling factor, $\alpha_1, \alpha_2$ as the excitation background.

Substituting (3) into (1) yields

$$\begin{aligned}&\rho_1\tau_t U_\tau + \rho_1\xi_x^3 U_{\xi\xi\xi} - 3\rho_1\rho_2\xi_x(UV)_\xi + (\rho_{1t} - 3\rho_1\alpha_{2x})U + \rho_1(\xi_t - 3\alpha_2\xi_x + \xi_{xxx})U_\xi \\ &+ 3\rho_1\xi_x\xi_{xx}U_{\xi\xi} - 3\rho_2\alpha_{1x}V - 3\rho_2\alpha_1\xi_x V_\xi + \alpha_{1t} + \alpha_{1xxx} - 3(\alpha_1\alpha_2)_x = 0, \\ &\rho_1\xi_x U_\xi + \alpha_{1x} = \rho_2\eta_y V_\eta + \alpha_{2y},\end{aligned} \tag{4}$$

Requiring $U(\xi,\eta,t)$ to satisfy (2) and $u(x,y,t)$ to be a solution of (1), one get the set of equations

$$\xi_{xx} = 0, \alpha_1 = 0, \alpha_{2y} = 0, \rho_{1t} - 3\rho_1\alpha_{2x} = 0, \rho_1\xi_x = \rho_2\eta_y, \tau_t = \xi_x^3 = \rho_2\xi_x. \tag{5}$$

It can be inferred from (5)

$$\xi(x,t) = \kappa(t)x + \omega(t), \eta(y,t) = \iota(t)y + \upsilon(t), \tag{6}$$

where $\kappa = \kappa(t), \iota = \iota(t), \omega = \omega(t), \upsilon = \upsilon(t)$ are some undetermined functions of the specified variable $t$. Substituting (6) into (5) and after some algebra, one can find out

$$\rho_1 = \rho_2 = \kappa^2, \xi = \kappa x + w, \eta = \kappa x + \upsilon, \tau = \int^t \kappa^3 dt, \alpha_1 = 0, \alpha_2 = \frac{1}{3\kappa}(\kappa_t x + \omega_t). \tag{7}$$

Combining (2) and (7), SMT can be described as

$$\begin{aligned}u(x,y,t) &= \kappa^2 U(\xi,\eta,\tau), \\ v(x,y,t) &= \kappa^2 V(\xi,\eta,\tau) + \frac{1}{3\kappa}(\kappa_t x + \omega_t),\end{aligned} \tag{8}$$

with

$$\begin{aligned}\xi(x,t) &= \kappa(t)x + \omega(t), \\ \eta(y,t) &= \kappa(t)x + \upsilon(t), \\ \tau(t) &= \int^t \kappa^3 dt,\end{aligned} \tag{9}$$

where $\kappa(t), \omega(t), \upsilon(t)$ are three arbitrary functions of time $t$.

It is worth emphasizing that SMT (8) with self-similar variables expressions (9) are very important because it enables self mapping of solutions for the

(2+1)-dimensional KdV equation, which including not only solitons and periodic waves but also dromions and lumps as well as other arbitrary solutions.

In what follows we provide the main results of 2DRWs excited on zero-background for the (2+1)-dimensional KdV equation (1) based on SST (14) with (15). Without lose generality, and in order to avoid the singularity of the solution, here we do choose

$$\kappa = \frac{\kappa_0}{1+a^2 t^2}, \tag{10}$$

where $a>0$ and $\kappa_0 >0$ are positive free real parameters. Of course, other reasonable expressions can be chosen, as long as there are no singular points in the entire time region $\{-\infty, \infty\}$.

Substitution of (10) to (8) gives

$$\begin{aligned}
u(x,y,t) &= \frac{\kappa_0^2}{(1+a^2 t^2)^2} U(\xi,\eta,\tau), \\
v(x,y,t) &= \frac{\kappa_0^2}{(1+a^2 t^2)^2} V(\xi,\eta,\tau) - \frac{2a^2 t}{3(1+a^2 t^2)} x + \frac{1+a^2 t^2}{3\kappa_0} w_t,
\end{aligned} \tag{11}$$

with

$$\begin{aligned}
\xi(t) &= \frac{\kappa_0^2}{(1+a^2 t^2)^2} x + \omega(t), \\
\eta(t) &= \frac{\kappa_0^2}{(1+a^2 t^2)^2} y + \upsilon(t), \\
\tau(t) &= \tau_0 + \frac{\kappa_0^3}{8a}\left[3\arctan(at) + \frac{3at}{1+a^2 t^2} + \frac{2at}{(1+a^2 t^2)^2}\right].
\end{aligned} \tag{12}$$

In what follows we will give the three most basic forms of induced 2DRW excited on the zero background in (2+1)-dimensional KdV equation (1).

We first discuss the line-soliton solutions, which are a kinds of basic 2D solitary waves decaying exponentially in the $(x,y)$-plane except along certain rays in the (2+1)-dimensional nonlinear models. Based on SMT with (12) and using the line-soliton for the equation (3) derived in [17], we obtain the line soliton induced 2DRWs of the equation (1) as follows

$$u(x,y,t) = -\frac{\kappa_0^2 k_1 l_1}{2(1+at^2)^2}\left[\text{sech}\left(\frac{1}{2}(k_1\xi+l_1\eta-k_1^3\tau)\right)\right]^2,$$

$$v(x,y,t) = -\frac{\kappa_0^2 k_1^2}{2(1+at^2)^2}\left[\text{sech}\left(\frac{1}{2}k_1\xi+l_1\eta-k_1^3\tau\right)\right] - \frac{2a^2 t}{3(1+a^2t^2)}x + \frac{1+a^2t^2}{3\kappa_0}\omega_\rho,$$

(13)

with the self-similarity variables expressions (9).

Figure 1 displays the line-soliton-induced 2DRWs of physical field $u$ in equation (1), the parameter selection is shown in Figure 1. We see that the line-soliton-induced 2DRW excited on zero-background in the $(x,y)$-plane and can be a very large amplitude wave at $t=0$ because the solution (13) exists an amplitude scaling factor $\frac{\kappa_0}{1+(at)^2}$. Apparently, such 2DRW is not only quickly decays but also can be not travel with time by the appropriate selection of arbitrary functions.

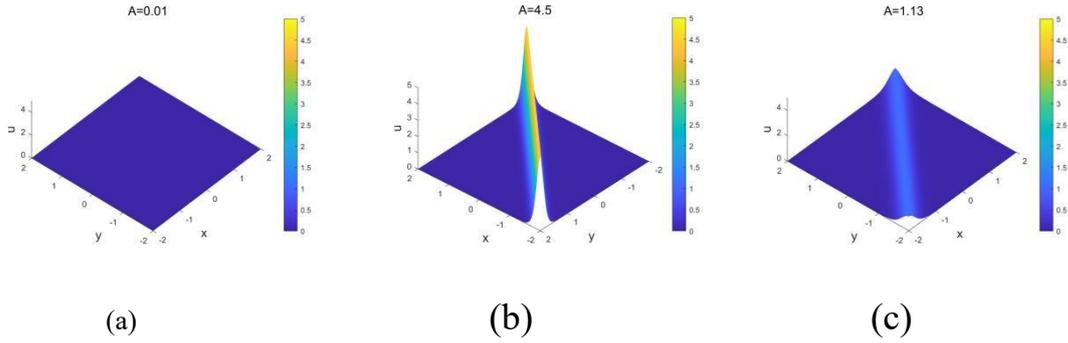

(a) (b) (c)

FIG. 1. (a)–(c) Profiles of the fundamental line-soliton-induced 2DRW of $u$ in the $(x,y)$ plane at (a) $t=-1$, (b) $t=0$ and (c) $t=0.2$. The parameters and the arbitrary functions are selected as selected as $a=5, k_1=2, l_1=-2, \kappa_0=3$ and $\omega=k_1^3\tau, \upsilon=0$, respectively.

We then discuss the dromion solutions, which are another kind of 2D solitons and are driven by both two perpendicular and non-perpendicular line or curved line ghost solitons. Based on SMT with (12) and the dromion for equation (3) derived in Ref.[17], we obtain the dromion-induced 2DRWs of the equation (1) as follows

$$u(x,y,t) = -\frac{2\kappa_0^2 k_1 l_1 (K-1)\exp\left[k_1\xi + l_1\eta - (k_1^3 + l_1^3)\tau\right]}{(1+a^2t^2)^2 \left\{1 + \exp(k_1\xi - k_1^3\tau) + \exp(l_1\eta - l_1^3\tau) + K\exp\left[k_1\xi + l_1\eta - (k_1^3 + l_1^3)\tau\right]\right\}^2},$$

$$v(x,y,t) = \frac{2\kappa_0^2 k_1^2 (K-1)\exp\left[k_1\xi + l_1\eta - (k_1^3 + l_1^3)\tau\right]}{(1+a^2t^2)^2 \left\{1 + \exp(k_1\xi - k_1^3\tau) + \exp(l_1\eta - l_1^3\tau) + K\exp\left[k_1\xi + l_1\eta - (k_1^3 + l_1^3)\tau\right]\right\}^2}$$
$$-\frac{2a^2 t}{3(1+a^2t^2)}x + \frac{1+a^2t^2}{3\kappa_0}\omega_t,$$

(14)

with the self-similarity variables expressions (9).

Again, we also see that the dromion-induced 2DRW of the physical field $u$ is excited on zero-background in the $(x,y)$-plane and is a very large amplitude wave at $t = 0$ from Figure 2. Apparently, such 2DRW is not only quickly decays but also can be not travel with time by the appropriate selection of arbitrary functions.

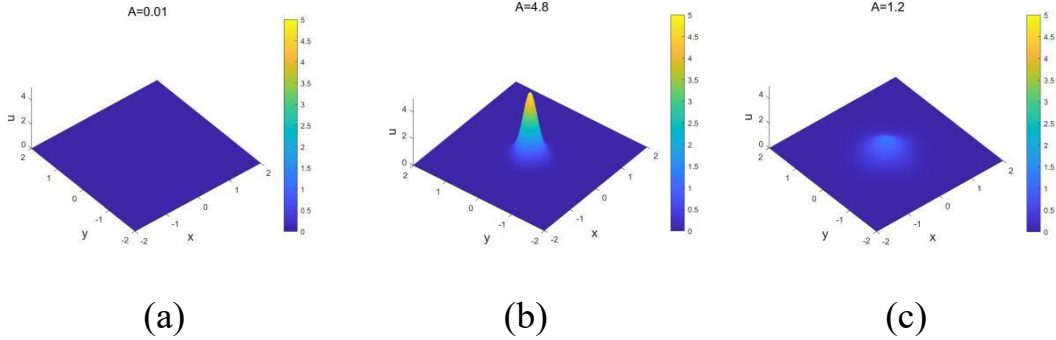

(a)　　　　　　　　　　(b)　　　　　　　　　　(c)

FIG.2. (a)–(c) Profiles of the fundamental dromion-induced 2DRW of $u$ in the $(x,y)$ plane at (a) $t = -1$, (b) $t = 0$ and (c) $t = 0.2$. The parameters and the arbitrary function are selected as $a = 5, k_1 = 2, l_1 = 2, \kappa_0 = 4, K = 0.4$ and $\omega = k_1^3\tau, \upsilon = l_1^3\tau$, respectively.

Next, we consider the lump solutions, which are third basic 2D solitons described by the rational solutions, are algebraically localized in two-dimensional space and travel in time. Lumps is localized. Based on SMT with (12) and using the lump of equation (3) derived in Ref. [18], we obtained the lump-induced 2DRWs of the equation (1) as follow

$$u(x,y,t) = \frac{8\kappa_0^2 k_2 \left[ l_1 l_2 (k_1\xi + l_1\eta)^2 + 2(l_2^2 - l_1^2)(k_1\xi + l_1\eta)(k_2\xi + l_2\eta) - l_1 l_2 (k_2\xi + l_2\eta)^2 \right]}{l_1 (1+a^2t^2)^2 \left[ A + (k_1\xi + l_1\eta)^2 + (k_2\xi + l_2\eta)^2 \right]^2},$$

$$v(x,y,t) = \frac{4\kappa_0^2 k_2}{l_1^2(1+a^2t^2)^2} \left[ \frac{(l_1^2 + l_2^2)}{A + (k_1\xi + l_1\eta)^2 + (k_2\xi + l_2\eta)^2} + \frac{(l_2(k_1\xi + l_1\eta) - l_1(k_2\xi + l_2\eta))}{\left[ A + (k_1\xi + l_1\eta)^2 + (k_2\xi + l_2\eta)^2 \right]^2} \right] \quad (15)$$

$$- \frac{2a^2 t}{3(1+a^2t^2)} x + \frac{1+a^2t^2}{3\kappa_0} \omega_t,$$

with the self-similarity variables expressions (9) and $A$ is an arbitrary constant.

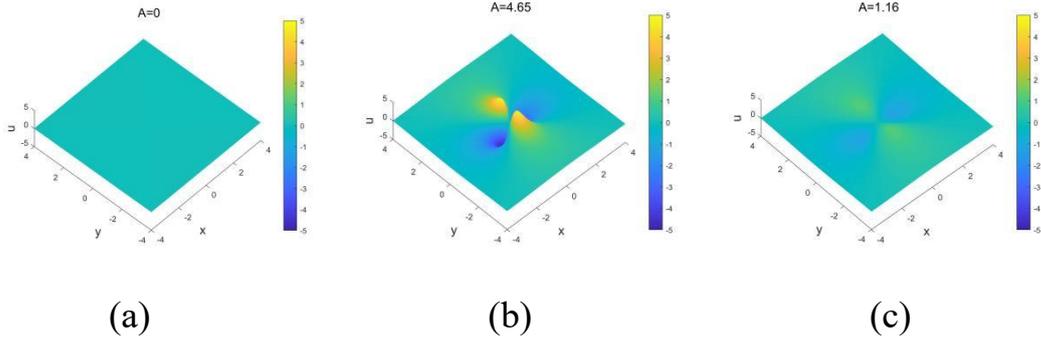

(a)              (b)              (c)

FIG.3. (a)–(c) Profiles of the fundamental lump-induced 2DRW of $u$ in the $(x,y)$ plane at (a) $t=-1$, (b) $t=0$ and (c) $t=0.2$. The parameters and the arbitrary functions are selected as $a=5, k_1=0.6, k_2=0.6, l_1=-1, l_2=-0.1, \kappa_0=2$ and $A=-0.5, \omega=0, \upsilon=0$, respectively.

Similarly, we also see that the lump-induced 2DRW of physical field $u$ is excited on zero- background in the $(x,y)$-plane and is a very large amplitude wave at $t=0$ in Figure 3. Because the solution (15) exists an amplitude scaling factor $\frac{\kappa_0}{1+(at)^2}$, such 2DRW quickly decays quickly with time.

In summary, we have proposed a self-mapping transformation with three arbitrary functions of time for the (2+1)-dimensional KdV equations (1). We have also constructed three kinds of the fundamental induced 2DRWs excited on zero-background in the equations (1). We expect that these findings may motivate research efforts towards the generation of the 2DRWs in hydrodynamics, nonlinear optics, plasma physics, Bose-Einstein condensation dynamics and so on. In addition, the typical KP equation and DS equation have also been made some progress, the related work will be reported separately. So we can expect to pave a new path for generate 2DRWs excited on zero-background and deeply study the dynamic properties of their interaction .

**Acknowledgments:** Our advance to this field have benefited from valuable collaborations and discussions with numerous colleagues and friends, in particular Prof. Da-Jun Zhang, Doctor Lei Wu and Doctor Zhao Zhang. Jie-fang Zhang extends further thanks to Prof. Wen-xiu Ma and Prof. Zhi-jun Qiao for sparking his interest in the wider aspects of the self-similarity and the rogue wave when he visited the United States.

The work was supported by the National Natural Science Foundation of China (Grant Nos. 61877053).